\documentclass[aps,graphicx,twocolumn]{revtex4}
\usepackage{graphicx}% Include figure files
\usepackage{color}
\usepackage{amsmath}
\usepackage{amsmath}
\usepackage{graphicx}
\usepackage{subfigure}
\usepackage{textcomp}
\usepackage{color}

\begin{document}

\title{Experimental one-step deterministic entanglement purification}

\author{Cen-Xiao Huang$^{1,4}$\footnote{These two authors contributed equally to this work}, Xiao-Min Hu$^{1,4}$\footnote{These two authors contributed equally to this work}, Bi-Heng Liu,$^{1,4}$\footnote{bhliu@ustc.edu.cn} Lan Zhou,$^{3,5}$ Yu-Bo Sheng,$^{2,5}$\footnote{shengyb@njupt.edu.cn}, Chuan-Feng Li,$^{1,4}$\footnote{cfli@ustc.edu.cn} Guang-Can Guo$^{1,4,5}$}
\address{$^1$CAS Key Laboratory of Quantum Information, University of Science and Technology of China,
Hefei 230026, China\\
$^2$College of Electronic and Optical Engineering \& College of Microelectronics, Nanjing University of Posts and Telecommunications, Nanjing 210003, China\\
$^3$School of Science, Nanjing University of Posts and Telecommunications, Nanjing, 210003, China\\
$^4$Synergetic Innovation Center of Quantum Information and Quantum Physics, University of Science and Technology of China, Hefei 230026, China\\
 $^5$Institute of Quantum Information and Technology, Nanjing University of Posts and Telecommunications, Nanjing, 210003, China\\}

%\date{\today }
\begin{abstract}
Entanglement purification is to distill high-quality entangled states from low-quality entangled states. It is a key step in quantum repeaters, determines the efficiency and communication rates of quantum communication protocols, and is hence of central importance in long-distance communications and quantum networks. In this work, we report the first experimental demonstration of deterministic entanglement purification using polarization and spatial mode hyperentanglement.  After purification, the fidelity of polarization entanglement arises from $0.268\pm0.002$ to $0.989\pm0.001$. Assisted with robust spatial mode entanglement, the total purification efficiency can be estimated as  $10^{9}$ times that of  the entanglement purification protocols using two copies of entangled states when one uses the spontaneous parametric down-conversion sources. Our work may have the potential to be implemented as a part of  full repeater protocols.

\end{abstract}

\date{\today }

\maketitle

\section{Introduction}
Entanglement plays an important role in quantum communications and quantum computations. Quantum teleportation~\cite{teleportation}, entanglement-based quantum key distribution~\cite{QKD,QKD1,QKD2}, entanglement-based quantum secure direct communication~\cite{QSDC1,QSDC2,QSDC3,QSDC4,QSDC5,QSDC6},  distributed quantum computation~\cite{distributed} all require to share nonlocal maximally entangled states. However, during the transmission, the maximally entangled states will degrade into mixed states. Such polluted entanglements will decrease the communication efficiency and even make the quantum communication become insecure.

Entanglement purification is a powerful tool to distill high-quality entangled states from the polluted low-quality entangled states~\cite{purification1}, and it is a key step in quantum repeaters~\cite{repeater}.  Entanglement purification was first proposed by Bennett et al in 1996 \cite{purification1} and was well developed in both theory and experiment. Existing entanglement purification protocols can be divided into two groups. The first group is the probabilistic entanglement purification \cite{purification2,purification3,purification4,purification5,purification6,purification7,purification8,purification9,purificationadd1,purification10,purification11,purification12,purification13,purification14,purification15,purification16}. In these protocols, two nonlocal parties require to use two identical noisy pairs of entangled states. After performing the controlled-not~(CNOT) operations or other similar operations, they measure the second copy to decide whether the purification is successful or not.
If the purification is successful, they will obtain a high-fidelity entangled pair with some probability. Otherwise, both pairs of entangled states should be discarded. This kind of entanglement purification protocols can only purify bit-flip errors in a round. If there exits phase-flip errors, the parties can purify the phase-flip errors in a next round by transforming the phase-flip errors to the bit-flip errors with the Hadamard operations. Therefore, to obtain higher-fidelity entanglements, the nonlocal parties need to repeat these processes by sacrificing a large amount of low-fidelity entangled pairs. Such entanglement purification has been realized in linear optical systems~\cite{purificationexperiment1,purificationexperiment2}, atoms~\cite{purificationexperiment3}, and spins~\cite{purificationexperiment4}. Meanwhile, probabilistic entanglement purification protocols can also be realized using only one copy of  imperfect hyperentangled states~\cite{Hu2021,purificationexperiment6}. By constructing the CNOT gate between different degrees of freedom (DOFs) of single photons, i.e., the polarization-spatial mode and the polarization-time-bin, the imperfect spatial mode (time-bin) entanglement can be used to purify the polarization entanglement. The probabilistic entanglement purification was also extended to use two non-identical pairs. It was shown that the discarded pairs still have entanglement and can be reused in the next purification round, if the purification is a failure~\cite{purification14}.

Another group is the deterministic entanglement purification. It also uses hyperentanglements~\cite{zeilinger,determin1,determin2,determin3,determin4,determin5}. In some special noisy environments, if entangled states in spatial mode, time-bin,  frequency or other DOFs are more robust than that in the polarization DOF \cite{robust1,robust2,robust3}, the robust entanglement can be used to purify the fragile polarization entanglement deterministically. Deterministic entanglement purification was first used to purify the bit-flip error~\cite{determin1}. After performing the purification operation, the bit-flip error in polarization entanglement can be corrected completely with a success probability of 100\%. However, after purification, the robust spatial entanglement is consumed and the hyperentanglement becomes a high-fidelity polarization entanglement. In order to purify the phase-flip error in the next round, the parties still need to exploit the probabilistic entanglement purification protocol. In 2010, the deterministic polarization entanglement purification for both bit-flip error and phase-flip error was proposed~\cite{determin2}. They use the spatial entanglement to purify the bit-flip error and use the frequency entanglement to purify phase-flip error completely. However, this protocol requires the polarization-spatial-frequency hyperentanglement and the cross-Kerr nonlinearity to construct the quantum nondemolition measurement, which makes this protocol hard to realize using existing technologies. In the same year, the deterministic entanglement purification protocol using only polarization-spatial hyperentanglement was proposed~\cite{determin3,determin4}. In these protocols, both bit-flip error and phase-flip error in polarization DOF can be corrected deterministically using spatial entanglement in feasible linear optics. Deterministic entanglement purification was also used in double-server blind quantum computation in noisy environments~\cite{determin5}.

It is known that entanglement purification plays a key role in quantum repeaters, for it determines the efficiency and communication rates of quantum communication protocols, and is hence of central importance in the long-distance communications and quantum networks~\cite{purification12}. The probabilistic entanglement purification requires to perform the protocol for several rounds to obtain a high-fidelity entangled state, by sacrificing a large amount of low-fidelity entangled pairs. Moreover, these protocols require the fidelity of initial low-quality entangled state to be greater than 1/2. In this paper, based on  Ref.~\cite{determin3}, we report the first experiment of one-step deterministic polarization entanglement purification using hyperentanglement. This paper is organized as follows. In Section~II, we briefly introduce the one-step deterministic entanglement purification. In Section~III, we explain the experiment setup and results. In Section~IV, we make a discussion and conclusion.

\begin{figure}
\includegraphics[width=9cm]{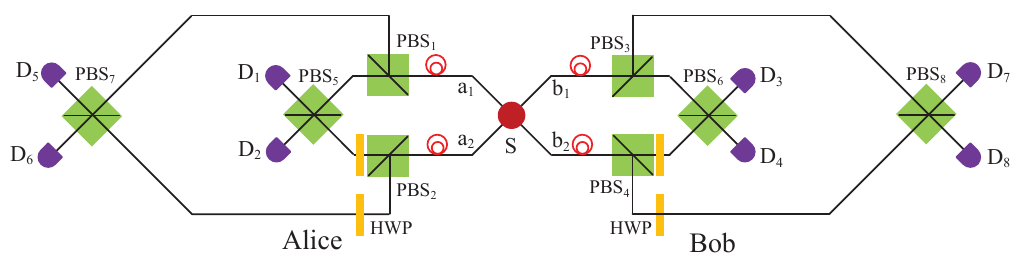}
\caption{Protocol of the deterministic entanglement purification using hyperentanglement~\cite{determin3}. PBS is the polarizing beam splitter. HWP is the half-wave plate which can convert the horizontally polarized photon to the vertically polarized photon, vice versa. $D_{i}(i=1,2,\cdots,8)$ is the single photon detector (SD).}
\label{fig1}
\end{figure}

\begin{figure*}[tph!]
\includegraphics[width=0.9\textwidth]{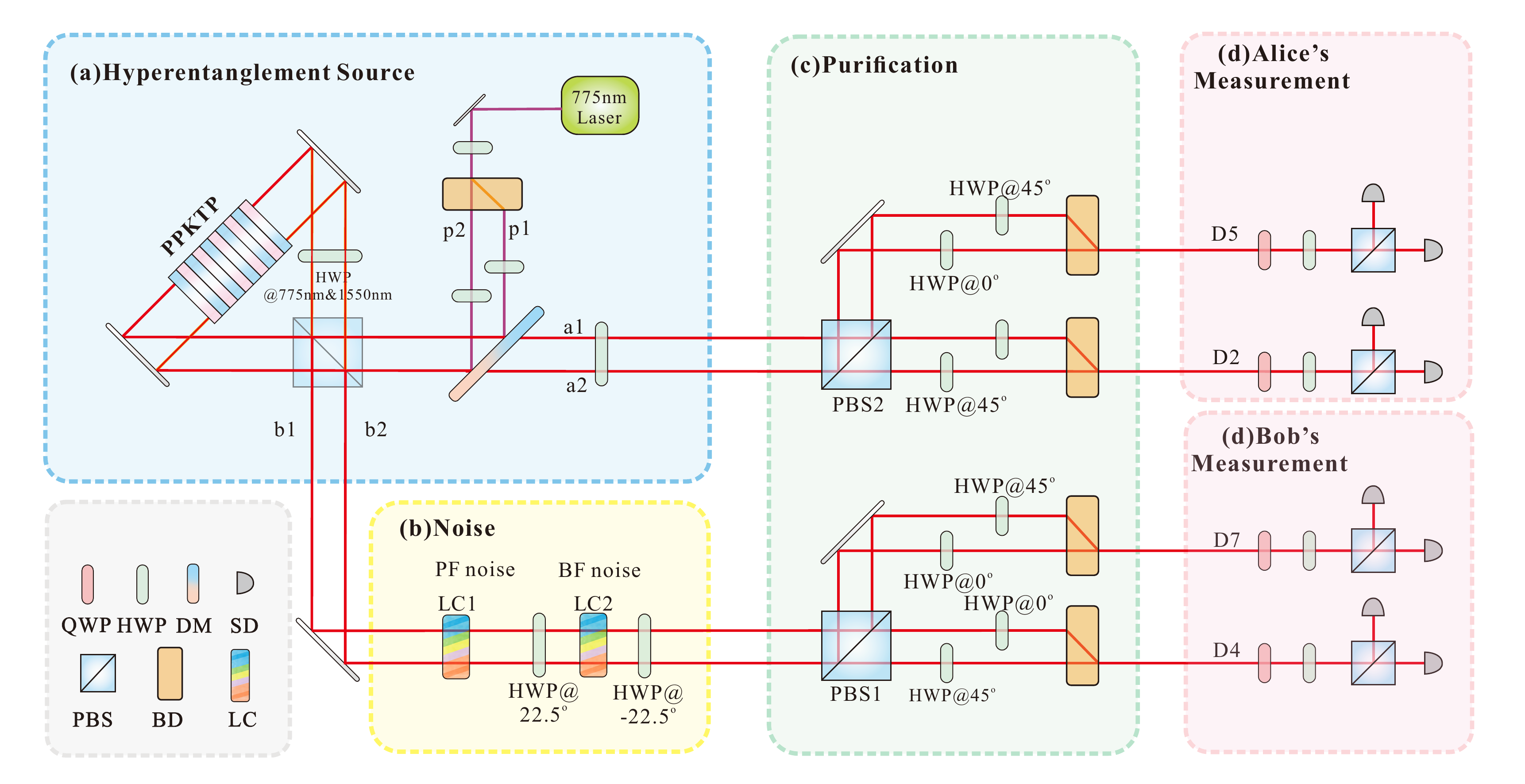}
\vspace{-0.5cm}
\caption{Experimental setup. (a) Preparation of hyperentangled states. The pump light beam (775~nm) is separated into two spatial modes by the beam displacer (BD). The two beams are injected into a Sagnac interferometer to pump a type-II $(\left|H\rangle_{775nm} \longrightarrow|H\rangle_{1550nm} \otimes|V\rangle_{1550nm}\right)$ cut periodically poled potassium titanyl phosphate (PPKTP) crystal and generate the two-photon polarization entanglement. It can finally generate the hyperentanglement $\frac{1}{2}(|H\rangle|H\rangle+|V\rangle|V\rangle)\otimes(|a_{1}\rangle|b_{1}\rangle+|a_{2}\rangle|b_{2}\rangle)$. DM represents the dichroic mirror. (b) Quantum noisy channel. The noise loading device consists of two HWPs and two liquid crystals (LCs). By controlling the voltages loaded on LCs, we can load any bit-flip (BF) noise and white noise. In this experiment, we loaded 30\%, 50\% and 70\% noise for each type of noise, respectively (details see Appendix). (c) Entanglement purification process. The setup consists of  HWPs and PBSs. Through the postselection of
PBS1, PBS2, and HWPs, we can purify polarization entanglement. We can obtain $|\phi^{+}\rangle_{P}$ in output modes $D_{2}D_{4}$, $D_{5}D_{7}$, $D_{2}D_{7}$ or $D_{4}D_{5}$. (d) Polarization state tomography setup. QWP represents the quarter-wave plate.}
\label{fig2}
\end{figure*}

\section{Basic principle of deterministic entanglement purification}
Now we first briefly describe the basic principle of such deterministic entanglement purification~\cite{determin3}. As shown in Fig.~\ref{fig1}, the hyperentanglement source emits  one pair of  hyperentangled state of the form
\begin{eqnarray}
|\phi\rangle&=&|\phi^{+}\rangle_{P}\otimes|\phi^{+}\rangle_{S}\nonumber\\
&=&\frac{1}{2}(|H\rangle|H\rangle+|V\rangle|V\rangle)\otimes(|a_{1}\rangle|b_{1}\rangle+|a_{2}\rangle|b_{2}\rangle).
\end{eqnarray}
After the photon pair transmits through a noisy channel, the polarization part of the hyperentangled state becomes a mixed state
\begin{eqnarray}
\rho_{P}&=&F_{1}|\phi^{+}\rangle_{P}{}_{P}\langle\phi^{+}|+F_{2}|\phi^{-}\rangle_{P}{}_{P}\langle\phi^{-}|\nonumber\\
&+&F_{3}|\psi^{+}\rangle_{P}{}_{P}\langle\psi^{+}|
+F_{4}|\psi^{-}\rangle_{P}{}_{P}\langle\psi^{-}|.
 \end{eqnarray}
Here
\begin{eqnarray}
|\phi^{-}\rangle_{P}=\frac{1}{\sqrt{2}}(|H\rangle|H\rangle-|V\rangle|V\rangle),\nonumber\\
|\psi^{\pm}\rangle_{P}=\frac{1}{\sqrt{2}}(|H\rangle|V\rangle\pm|V\rangle|H\rangle).
\end{eqnarray}
$a_{1}$, $a_{2}$, $b_{1}$, and $b_{2}$ are the spatial modes as shown in Fig.~\ref{fig1}. Suppose that the entanglement in the other DOF, i.e. spatial mode, is robust. Therefore, the whole mixed state can be written as
\begin{eqnarray}
\rho=\rho_{P}\otimes\rho_{S},
\end{eqnarray}
with $\rho_{S}=|\phi^{+}\rangle\langle\phi^{+}|$.
The whole mixed state can be described as follows. It is in the state $|\phi^{+}\rangle_{P}\otimes|\phi^{+}\rangle_{S}$ with the probability of $F_{1}$, in the state $|\phi^{-}\rangle_{P}\otimes|\phi^{+}\rangle_{S}$ with the probability of $F_{2}$, and so on. As shown in Fig.~\ref{fig1}, the purification process will make the state $|\phi^{+}\rangle_{P}\otimes|\phi^{+}\rangle_{S}$ and  $|\phi^{-}\rangle_{P}\otimes|\phi^{+}\rangle_{S}$  become $|\phi^{+}\rangle_{P}$ in the output modes $D_{2}D_{4}$ or $D_{5}D_{7}$. On the other hand, $|\psi^{+}\rangle_{P}\otimes|\phi^{+}\rangle_{S}$ and  $|\psi^{-}\rangle_{P}\otimes|\phi^{+}\rangle_{S}$ will become $|\psi^{+}\rangle_{P}$ in the output modes $D_{2}D_{7}$ or $D_{4}D_{5}$. By selecting the output modes $D_{2}D_{4}$, $D_{5}D_{7}$, $D_{2}D_{7}$ or $D_{4}D_{5}$, the nonlocal parties can obtain the state $|\phi^{+}\rangle_{P}$ or $|\psi^{+}\rangle_{P}$. If they obtain $|\psi^{+}\rangle_{P}$, by adding a bit-flip operation on one of the photons, they can convert $|\psi^{+}\rangle_{P}$ to $|\phi^{+}\rangle_{P}$ deterministically.

\begin{table*}[tph!]
\caption{Experiment result. The table shows the fidelity (corresponding to the state $|\phi^+ \rangle_{P(S)}$) of the states in different DOFs before and after purification. Each line represents the experimental results for one type of noise, e.g. BF0.7 represents 70\% bit-flip noise is loaded on the polarization DOF qubit, White0.3 represents 30\% white noise is loaded. 0.295(1) represents $0.295\pm0.001$. }
\centering
\begin{tabular}{c|cc|cccc}
\hline
\hline
      & \multicolumn{2}{c|}{Before purification}                                   & \multicolumn{4}{c}{After purification} \\ \hline
Noise    & \multicolumn{1}{c}{polarization} & \multicolumn{1}{c|}{ \quad path \quad\quad} & \quad D2D4\quad\quad  &\quad D5D7 \quad\quad&\quad D2D7\quad\quad &\quad D5D4\quad\quad \\ \hline
BF0.7    & 0.295(1) & 0.991(1) & 0.990(1) & 0.986(1) & 0.983(1) & 0.991(1)    \\ \hline
BF0.5    & 0.496(1) & 0.990(1) & 0.990(1) & 0.987(1) & 0.985(1) & 0.989(1)   \\ \hline
BF0.3    & 0.693(1) & 0.990(1) & 0.989(1) & 0.988(1) & 0.984(1) & 0.988(1)    \\ \hline
White0.7 & 0.268(2) & 0.990(1) & 0.989(1) & 0.985(1) & 0.980(1) & 0.983(1)   \\ \hline
White0.5 & 0.493(1) & 0.990(1) & 0.987(1) & 0.988(1) & 0.987(1) & 0.985(1)   \\ \hline
White0.3 & 0.651(1) & 0.990(1) & 0.988(1) & 0.984(1) & 0.983(1) & 0.983(1) \\ \hline \hline
\end{tabular}
\label{table1}
\end{table*}

\section{Experimental setup and results}
\par As shown in Fig.~\ref{fig2}(a), we use the hyperentangled state generated by a type-II cut PPKTP crystal to demonstrate the entanglement purification. The continuous-wave (CW) laser at 775~nm is separated into two spatial modes (p1 and p2) by a beam displacer (BD) and then injected to a polarization Sagnac interferometer to generate the hyperentangled photon pairs. The final state is the superposition of the state in each mode. Thus, we can generate the hyperentanglement $|\phi\rangle=|\phi^+\rangle_{P} \otimes |\phi^+\rangle_{S} $ by tuning the relative phase between the two spatial modes~\cite{Hu2018,Hu2021}. We used 120~mW pumped light to excite 2400 photon pairs per second. The coincidence efficiency of the entangled source is 20\%.

In the one-step deterministic purification protocol, we can correct both bit-flip and phase-flip errors in polarization entanglement. As shown in Fig.~\ref{fig2}(b), we introduce extra noises in the polarization DOF, which include two types of noise --- bit-flip (BF) noise ($\rho_{P}=|\phi^+\rangle_{P}{}_{P}\langle\phi^+|\rightarrow F|\phi^+\rangle_{P}{}_{P}\langle\phi^+|+(1-F)|\psi^+\rangle_{P}{}_{P}\langle\psi^+| $) and white noise ($\rho_{P}=|\phi^+\rangle_{P}{}_{P}\langle\phi^+| \rightarrow F|\phi^+\rangle_{P}{}_{P}\langle\phi^+|+\frac{1-F}{3}|\phi^-\rangle_{P}{}_{P}\langle\phi^-| +\frac{1-F}{3}|\psi^+\rangle_{P}{}_{P}\langle\psi^+|+\frac{1-F}{3}|\psi^-\rangle_{P}{}_{P}\langle\psi^-|$). We use two HWPs (setting at $22.5^{\circ}$ and $-22.5^{\circ}$) and a liquid crystal ($LC2$ set at $0^{\circ}$) phase plate to add the BF noise in polarization DOF. When the voltage of $LC2$ is $V_{\pi}$, BF ($\sigma_x=|H\rangle\langle V|+|V\rangle\langle H|$) operation is performed. On the other hand, the polarization stays the same when voltage $V_0$ is applied to the $LC2$. For white noise, we need an extra $LC1$ (setting at $0^{\circ}$) to perform the phase-flip operation ($\sigma_z=|H\rangle\langle H|-|V\rangle\langle V|$). By changing the LCs' voltages periodically, we can tune the ratio of noise from 0 to 1. In our experiment, we loaded 30\%, 50\% and 70\% noise for each type of noise, respectively (details see Appendix). The experimental fidelity of the states before purification are listed in Table~\ref{table1}. The average fidelity of the entanglement in spatial mode DOF is about 0.99.

The process of polarization entanglement purification is shown in Fig.~\ref{fig2}(c). We use the polarizing beam splitter (PBS) to postselect the photons with different polarization, then combine the different spatial modes by a BD and two HWPs. Here we should mention that the experiment protocol in Fig.~\ref{fig2} is a little different from Fig.~\ref{fig1}, and the two HWPs of $45^{\circ}$ are set at the same spatial mode. Such a change will make the photons detected by coincidence outputs \{$D_2, D_4$\},\{$D_5, D_7$\}, \{$D_2, D_7$\} or \{$D_4, D_5$\} are in the state $|\phi^+\rangle_P$. All the photons after purification are nearly in the pure state $|\phi^+\rangle_P$, and we do not need extra operations on different outputs to get the same maximally entangled state.

After purification, the fidelities of states in polarization DOF will increase. For example, by loading 70\% white noise on the polarization part, the fidelity of the polarization entanglement is  $F=0.268\pm0.002$, and the fidelity of the spatial mode entanglement is  $F=0.990\pm0.001$ before purification.
After purification, the fidelity of polarization entanglement ($|\phi^+\rangle_{P}$) is $F_{D_2D_4}=0.989\pm0.001$, $F_{D_5D_7}=0.985\pm0.001$, $F_{D_2D_7}=0.980\pm0.001$, and $F_{D_4D_5}=0.983\pm0.001$, respectively.
On the other hand, by loading 30\% bit-flip noise, the fidelity of the polarization entanglement is $F=0.693\pm0.001$ and the fidelity of spatial mode entanglement is $0.990\pm0.001$. After  purification, the fidelity of polarization entanglement ($|\phi^+\rangle_{P}$) is $F_{D_2D_4}=0.989\pm0.001$, $F_{D_5D_7}=0.988\pm0.001$, $F_{D_2D_7}=0.984\pm0.001$, and $F_{D_4D_5}=0.988\pm0.001$, respectively. We show all the experiment results after purification in Table~\ref{table1} (details see Appendix).

\section{Discussion and conclusion}
We demonstrate the first deterministic entanglement purification using spatial mode entanglement~\cite{determin3}. Compared with all two-copy entanglement purification protocols, this experiment use only one pair of hyperentangled state. It will make the purification more efficient. For example, when one loads 20\% white noise, we only need to perform one step purification to obtain nearly perfect maximally entangled states. However, by using the purification protocol in Ref.~\cite{purification1}, we should perform the purification process for three times to obtain an entangled state with a fidelity of 0.905. Moreover, if we consider the spontaneous parametric down-conversion (SPDC) source, this deterministic entanglement purification becomes more efficient. By using one pair of hyperentangled state, the success probability of obtaining nearly perfect maximally entangled state is about $P_{s}\approx 0.001$~\cite{Hu2021}. By using two-copy entanglement purification~\cite{purification1}, in order to perform one bit-flip error purification and one phase-flip error purification, they should at least choose four copies of low-fidelity mixed states, with the probability of  $P^{4}_{s}\approx 10^{-12}$. Therefore, the whole purification efficiency can be estimated as $10^{9}$ times that of the entanglement purification protocols using two copies of entangled states with SPDC sources~\cite{purification1}.

On the other hand, Refs.~\cite{Hu2021,purificationexperiment6} also demonstrated the polarization entanglement purification using spatial mode entanglement and time-bin entanglement. In their protocols, the spatial mode entanglement~\cite{Hu2021} and time-bin entanglement~\cite{purificationexperiment6} can be used to purify the bit-flip error. After performing the purification successfully, such spatial mode entanglement and time-bin entanglement is consumed. In order to purify the phase-flip error, they should first perform the Hadamard operation to transform the phase-flip error to bit-flip error, and then select two copies of polarization states and perform the purification operation in a next round. In a practical transmission, the environment noise will cause the entanglement encoded in different DOFs result in different effect. In the entanglement purification protocols~\cite{Hu2021,purificationexperiment6}, they require that the fidelity of  polarization entanglement, spatial mode entanglement or time-bin entanglement are all larger than 1/2. If the noise makes the fidelity of  polarization entanglement degrade to less than 1/2, while the spatial mode entanglement or time-bin entanglement is still robust, using the approaches in Refs.~\cite{Hu2021,purificationexperiment6} may not obtain a higher-fidelity polarization entanglement. In this way, it is better to use the existing deterministic entanglement purification protocol.

In conclusion, we have reported the first experimental demonstration of the deterministic entanglement purification assisting with polarization and spatial mode hyperentanglement. We show the robust entanglement encoded in the other DOFs have the ability to purify the polarization entanglement deterministically. This work is general and can be effectively extended to use the robust entanglement in other DOFs, such as the time-bin, frequency, and so on. It also has the potential to be implemented as a part of full repeater protocols.\\
\textit{We note that a similar work was reported recently~\cite{Ecker2021}. }

\section{Appendix}
\subsection{Noise loading setup}
\par The noise loading setup is shown in Fig.~\ref{fig2}(b). LC1 is used to load PF noise ($\sigma_z$ operation) in polarization DOF. LC2 and two HWPs (set at $22.5^{\circ}$ and $-22.5^{\circ}$) are used to load BF noise ($\sigma_x$ operation) in polarization DOF. When the voltage of the $LC2$ phase plate is set at $V_0$, the polarization of the photon remains the same. When the voltage is $V_{\pi}$, it will introduce an extra $e^{i\pi}$ phase between $|H\rangle$ and $|V\rangle$. Combining the two HWPs, the $\sigma_x$ operation of polarization can be completed. The detailed formular derivation process can be written as
\begin{eqnarray}
|H\rangle &\xrightarrow{HWP_{22.5^{\circ}}}&\frac{1}{\sqrt{2}}(|H\rangle+|V\rangle)\\\nonumber
&\xrightarrow{LC_{V_{\pi}}}&\frac{1}{\sqrt{2}} (|H\rangle-|V\rangle)  \xrightarrow{HWP_{22.5^{\circ}}} |V\rangle,\\\nonumber
|V\rangle  &\xrightarrow{HWP_{22.5^{\circ}}}&\frac{1}{\sqrt{2}}(|H\rangle-|V\rangle)\\\nonumber
&\xrightarrow{LC_{V_{\pi}}}&\frac{1}{\sqrt{2}} (|H\rangle+|V\rangle)  \xrightarrow{HWP_{22.5^{\circ}}}|H\rangle.
\end{eqnarray}

For loading white noise, we control the proportion of loading $\sigma_z$ operation and $\sigma_x$ operation. By loading $\sigma_z$ and $\sigma_x$ operations, we can transform the state $|\phi^+\rangle_{P}$ to arbitrary Bell state as
$$|\phi^+\rangle_{P} \xrightarrow{\sigma_z}|\phi^-\rangle_{P},  |\phi^+\rangle_{P} \xrightarrow{\sigma_x}|\psi^+\rangle_{P}, |\phi^+\rangle_{P} \xrightarrow{\sigma_z\cdot\sigma_x}|\psi^-\rangle_{P}.$$

By adjusting the duty cycle between different voltages, any proportion of BF noise and white noise can be loaded. As shown in Fig. 3 and  Table~\ref{table2}, we take 10~s as a cycle. Various voltages are applied at different times in each cycle. In our experiment, we load $30\%$, $50\%$, and $70\%$ noise, respectively.

\begin{figure}[tph!]
\centering
\includegraphics[width=0.5\textwidth]{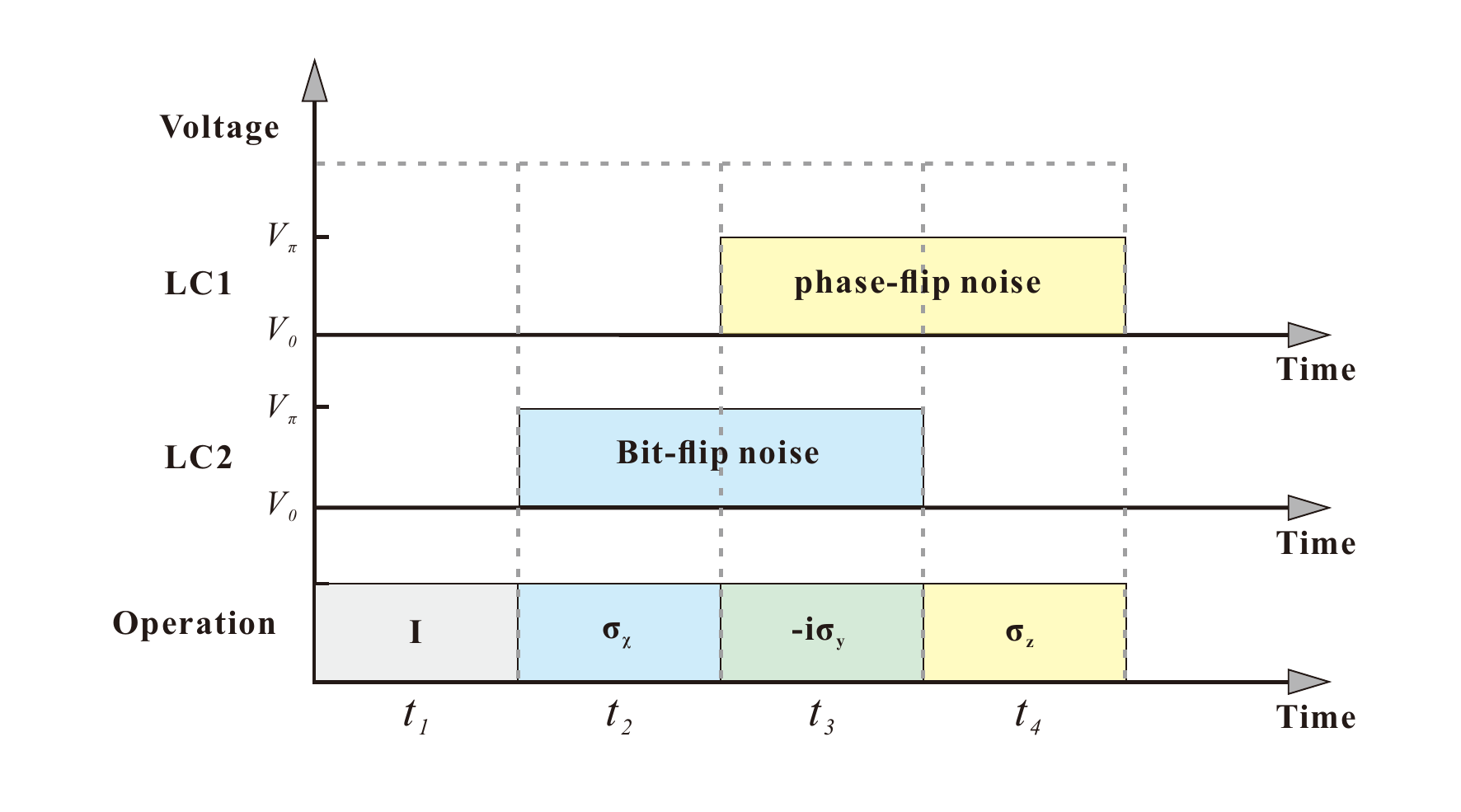}
\caption{Voltage timing diagram of two LC phase plates. For bit-flip noise, only the operations $I$ and $\sigma_x$ are needed. The voltage of LC1 is set at $V_0$, and by changing the ratio of $t_1$ and $t_2$, we can load any proportion of bit-flip noise. For white noise, both LCs are needed to realize the $\{I,\sigma_x,\sigma_y,\sigma_z\}$ operations (corresponding to time $t_1,t_2,t_3$ and $t_4$, respectively). With the help of these operations, we can transform $|\phi^+\rangle_P$ state to any other three Bell state. If the four Bell states occur alternately with the same probability, white noise is loaded.
The details of the time sequence in our experiment is listed in Table~\ref{table2}. }
\label{fig4}
\end{figure}

\begin{table}[tph!]
\caption{Time sequence of loading different noises. For bit-flip noise, only $LC2$ is needed, and the voltage of $LC1$ is set at $V_0$. For white noise, both $LC1$ and $LC2$ are needed, by setting the time sequence we get different ratios of noise.}
\centering
\begin{tabular}{c|c|c|c|c|c|c}
\hline \hline
Time & BF0.7 & BF0.5 & BF0.3 & white0.7 & white0.5 & white0.3               \\ \hline
\multicolumn{1}{c|}{$t_1$} & 3s  & 5s  & 7s  & 3s        & 5s      &  \multicolumn{1}{c}{7s} \\ \hline
\multicolumn{1}{c|}{$t_2$} & 7s  & 5s & 3s & 2.3s        & 1.67s      &  \multicolumn{1}{c}{1s} \\ \hline
\multicolumn{1}{c|}{$t_3$} & 0s  & 0s  & 0s   & 2.3s  & 1.67s & \multicolumn{1}{c}{1s} \\ \hline
\multicolumn{1}{c|}{$t_4$} & 0s  & 0s  & 0s  & 2.3s    & 1.67s  &  \multicolumn{1}{c}{1s} \\ \hline \hline
\end{tabular}
\label{table2}
\end{table}

\subsection{The measurement setup of spatial qubit}

\begin{figure}[tph!]
\centering
\includegraphics[width=0.5\textwidth]{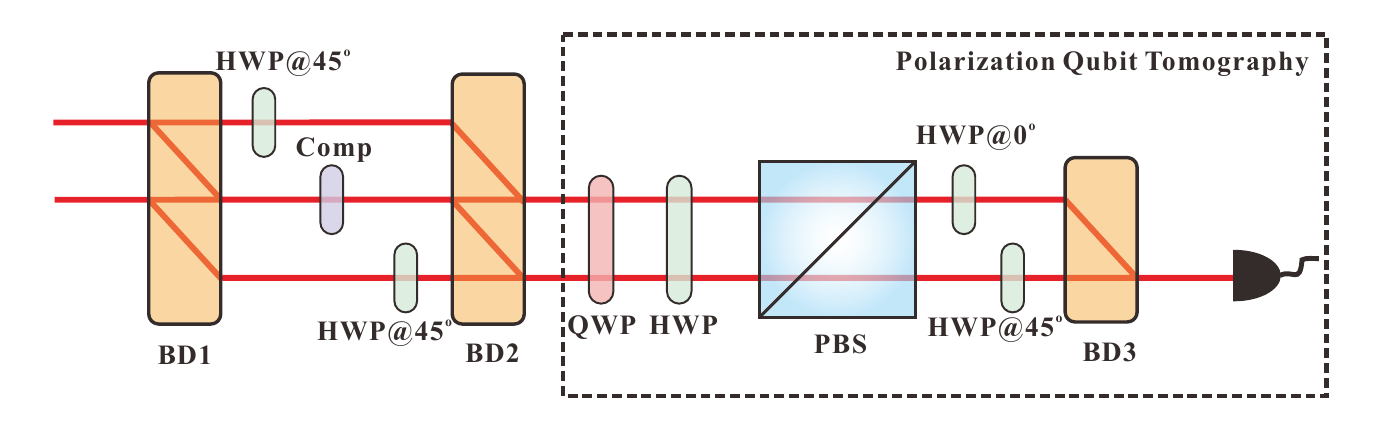}
\caption{Measurement of spatial qubit. BD1 and BD2 are used to convert the polarization and spatial qubit. The compensator (Comp) is used to compensate for the optical path difference. Thus the standard polarization tomography setup is used to measure the spatial qubit.}
\label{fig3}
\end{figure}

We have designed an experimental setup to measure the entanglement in spatial mode DOF using polarizers (Fig.~\ref{fig3}). We use BDs, HWPs, a QWP and a PBS to realize the state tomography of spatial mode entanglement. Notice that the first two BDs and HWPs realize the conversion between polarization and spatial qubits. At this time, the spatial mode DOF quantum state is converted to the polarization DOF quantum state, and then we use the standard polarization tomography setup for measurement.

\subsection{Purification results}
In Fig. 5 and Fig. 6, we show the density matrices of the polarization state before (left) and after (right) the purification in the case of BF noise and white noise, respectively, It is obvious that the fidelity of the purified quantum state can be  improved significantly. The exactly values of the fidelity of $|\phi^+\rangle_P$ under different noise ratios are shown in Table I.

%\textcolor{red}{The total success probability is $P=(\frac{1}{2})^{n}\ast\prod(P_{n})\ast(P_{s})^{2n}$. Here $n$ is the number of %purification round. $(\frac{1}{2})^{n}$ means that in each round, two copies of mixed states should be selected and only at least one %pair is remained. $P_{n}$ is the purification success probability in each round. Suppose the initial fidelity of Werner state is %$0.8$, we need to purify $n=3$ rounds to obtain the fidelity $0.9045$ with  $P\approx 6.6\ast 10^{-20}$.}

\begin{figure*}[tph!]
\centering
\includegraphics[width=0.8\textwidth]{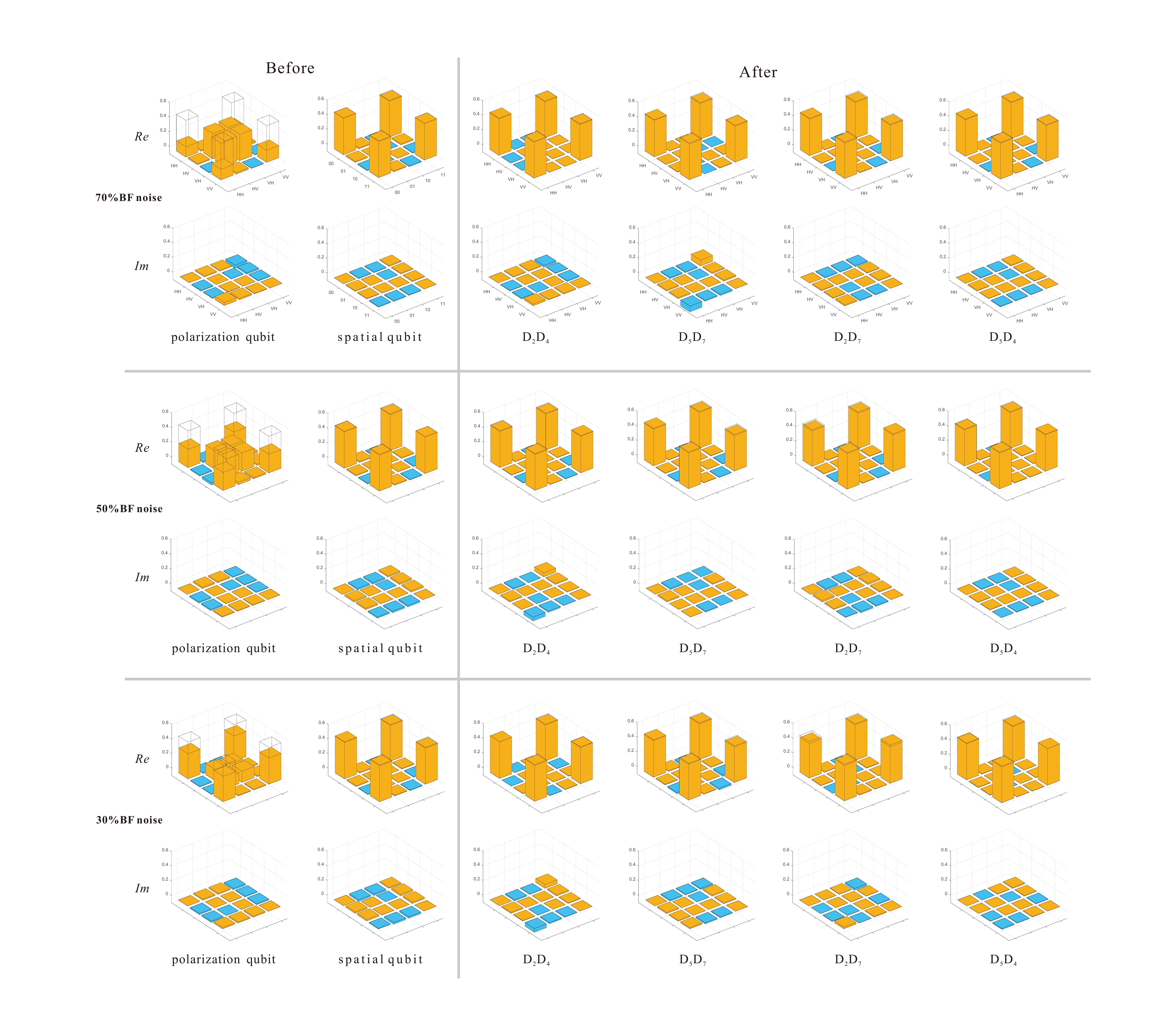}
\caption{The density matrices before and after purification in the case of BF noise. }
\label{fig5}
\end{figure*}

\begin{figure*}[tph!]
\centering
\includegraphics[width=0.8\textwidth]{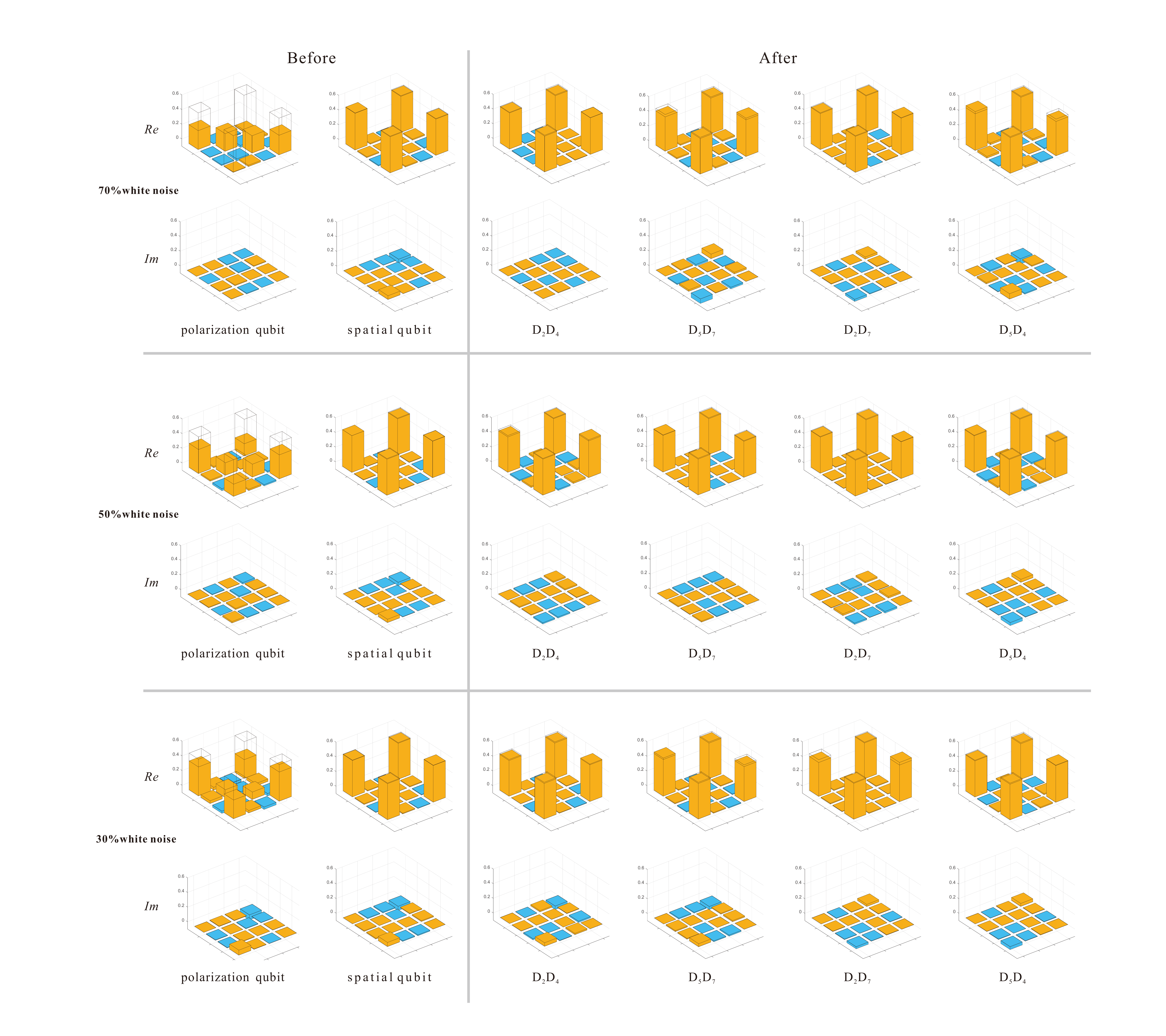}
\caption{The density matrices before and after purification in the case of white noise. }
\label{fig6}
\end{figure*}

\section*{ACKNOWLEDGEMENTS}
This work was supported by the National Key Research and Development Program of China (No.\ 2018YFA0306703, No.\ 2017YFA0304100), National Natural Science  Foundation of China (Nos. 11774335, 11734015, 11874345, 11821404, 11904357, 11974189, 12174367, 12175106), the Key Research Program of Frontier Sciences, CAS (No.\ QYZDY-SSW-SLH003), Science Foundation of the CAS (ZDRW-XH-2019-1), the Fundamental Research Funds for the Central Universities, USTC Tang Scholarship, Science and Technological Fund of Anhui Province for Outstanding Youth (2008085J02).


\begin{thebibliography}{99}

\bibitem{teleportation} Bennett CH, Brassard G, Crepeau C, et al. Teleporting an unknown quantum
state via dual classical and Einstein-Podolsky-Rosen channels. Phys Rev Lett
1993; 70: 1895-1899.

\bibitem{QKD} Ekert AK. Quantum cryptography based on Bell's theorem. Phys Rev Lett
1991; 67: 661-663.

\bibitem{QKD1}Kwek LC, Cao L, Luo W, et al. Chip-based quantum key distribution. AAPPS Bull 2021; 31:15.

\bibitem{QKD2}Hu XM, Zhang C, Guo Y, et al. Pathways for Entanglement-Based Quantum Communication in the Face of High Noise. Phys Rev Lett 127, 110505 (2021)

\bibitem{QSDC1} Long GL, Liu XS. Theoretically efficient high-capacity quantum key distribution
scheme. Phys Rev A 2002; 65: 032302.

\bibitem{QSDC2} Deng FG, Long GL, Liu XS. Two-step quantum direct communication protocol
using the Einstein-Podolsky-Rosen pair block. Phys Rev A 2003; 68: 042317.

\bibitem{QSDC3}Niu PH, Zhou ZR, Lin ZS, et al. Measurement-device-independent quantum communication without encryption. Sci Bull 2018; 63: 1345-1350.

\bibitem{QSDC4}Zhou L, Sheng YB, Long GL. Device-independent quantum secure direct communication against collective attacks. Sci Bull 2020; 65: 12-20.

\bibitem{QSDC5}Long GL, Zhang HR. Drastic increase of channel capacity in quantum secure direct communication using masking. Sci Bull 2021; 66: 1267-1269.

\bibitem{QSDC6}Qi Z, Li YH, Huang YW, et al. A 15-user quantum secure direct communication
network. Light Sc. Appl. 2021; 10:183.

\bibitem{distributed} Cirac JI, Ekert AK, Huelga SF, Macchiavello C. Distributed quantum computation over noisy channels. Phys Rev A 1999; 59: 4249.

\bibitem{purification1} Bennett CH, Brassard G, Popescu S, Schumacher B, Smolin JA, Wootters WK. Purification of noisy entanglement and faithful teleportation via noisy channels. Phys Rev Lett 1996; 76: 722.

\bibitem{repeater} Briegel HJ, D\"{u}r W, Cirac JI, Zoller P. Quantum repeaters: the role of imperfect local operations in quantum communication. Phys Rev Lett 1998; 81: 5932.

\bibitem{purification2} Deutsch D, Ekert A, Jozsa R, Macchiavello C, Popescu S, Sanpera A. Quantum privacy amplification and
the security of quantum cryptography over noisy channels. Phys Rev Lett 1996; 77: 2818-2821.

\bibitem{purification3} Murao M, Plenio MB, Popescu S, Vedral V, Knight PL. Multiparticle entanglement purification protocols.
Phys Rev A 1998; 57: R4075-R4078.

\bibitem{purification4} D\"{u}r W, Aschauer H, Briegel HJ. Multiparticle entanglement purification for graph states. Phys Rev Lett 2003; 91: 107903.

\bibitem{purification5} Pan JW,  Simon C,  Zellinger A. Entanglement purification for quantum communication. Nature 2001; 410: 1067-1070.

\bibitem{purification6} Cheong YW, Lee SW, Lee J, Lee HW. Entanglement purification for high-dimensional multipartite
systems. Phys Rev A 2007; 76: 042314.

\bibitem{purification7}  Sheng YB, Deng FG, Zhou HY. Efficient polarization-entanglement purification based on parametric
down-conversion sources with cross-kerr nonlinearity. Phys Rev A 2008; 77: 042308.

\bibitem{purification8}  Wang C, Zhang Y, Jin GS. Entanglement purification and concentration of electron-spin entangled states
using quantum-dot spins in optical microcavities. Phys Rev A 2011; 84: 032307.

\bibitem{purification9} Sheng YB, Zhou L, Long GL. Hybrid entanglement purification for quantum repeaters. Phys Rev A 2013; 88:
022302.

\bibitem{purificationadd1} Zwerger M, Briegel HJ, D\"{u}r W. Universal and optimal error thresholds for measurement-based entanglement
purification. Phys Rev Lett 2013; 110: 260503.

\bibitem{purification10}  Ren BC, Du FF, Deng FG. Two-step hyperentanglement purification with the quantum-state-joining
method. Phys Rev A 2014; 90: 052309.

\bibitem{purification11} Zhang H, Liu Q, Xu XS, Xiong J, Alsaedi A, Hayat T, Deng FG. Polarization entanglement purification
of nonlocal microwave photons based on the cross-kerr effect in circuit qed. Phys Rev A 2017; 96: 052330.

\bibitem{purification12} Miguel-Ramiro J, D\"{u}r W. Efficient entanglement purification protocols for d-level systems. Phys Rev A 2018; 98:
042309.

\bibitem{purification13} Krastanov S, Albert VV, Jiang L. Optimized entanglement purification. Quantum 2019; 3: 123123.

\bibitem{purification14} Zhou L, Zhong W,  Sheng YB. Purification of the residual entanglement. Opt. Express 2020; 28: 2291.

\bibitem{purification15} Riera-Sa\`{a}bat F, Sekatski P, Pirker A, D\"{u}r W. Entanglement-assisted entanglement purification. Phys Rev Lett 2021; 127: 040502.

\bibitem{purification16} Riera-Sa\`{a}bat F, Sekatski P, Pirker A, D\"{u}r W. Entanglement purification by counting and locating errors with entangling measurements. Phys Rev A 2021; 104: 012419.

\bibitem{purificationexperiment1}  Pan JW, Gasparoni S, Ursin R, Weihs G, Zeilinger A. Experimental entanglement purification of arbitrary
unknown states. Nature 2003; 423: 417-422.

\bibitem{purificationexperiment2} Chen LK, Yong HL, Xu P, et al. Experimental nested purification for a linear optical quantum repeater. Nat Photon 2017; 11:
695-699.

\bibitem{purificationexperiment3} Reichle R, Leibfried D, Knill E, et al. Experimental purification of two-atom entanglement. Nature (London) 2006; 443: 838-841.

\bibitem{purificationexperiment4} Kalb N, Reiserer AA, Humphreys PC, et al. Entanglement distillation between solid-state quantum network nodes.
Science 2017; 356: 928-932.

\bibitem{Hu2021}Hu XM, Huang CX, Sheng YB, et al. Long-distance entanglement purification for quantum communication. Phys Rev Lett 2021; 126: 010503.

\bibitem{purificationexperiment6} Ecker S, PhiliSohr P, Bulla L, et al. Experimental single-copy entanglement distillation. Phys Rev Lett 2021; 127: 040506.

\bibitem{zeilinger} Erhard M, Krenn M, Zeilinger A. Advances in high-dimensional quantum entanglement. Nat Rev Phys 2020; 2: 365.


\bibitem{determin1} Simon C, Pan JW. Polarization entanglement purification using spatial entanglement. Phys Rev Lett 2002; 89: 257901.

\bibitem{determin2} Sheng YB, Deng FG. Deterministic entanglement purification and complete nonlocal bell-state analysis with
hyperentanglement. Phys Rev A 2010; 81: 032307.

\bibitem{determin3} Sheng YB, Deng FG. One-step deterministic polarization-entanglement purification using spatial entanglement.
Phys Rev A 2010; 82: 044305.

\bibitem{determin4} Li XH. Deterministic polarization-entanglement purification using spatial entanglement. Phys Rev A 2010; 82:
044304.

\bibitem{determin5} Sheng YB, Zhou L. Deterministic entanglement distillation for secure double-server blind quantum
computation. Sci Rep 2015; 5: 7815.

\bibitem{robust1} Merolla JM, Mazurenko Y, Goedgebuer JP, Rhodes WT. Single-photon interference in sidebands of phase-modulated light for quantum cryptography. Phys Rev Lett 1999; 82: 1656.

\bibitem{robust2}Marcikic I, de Riedmatten H, Tittel W, Zbinden H, Legre M, Gisin N. Distribution of time-bin entangled qubits over 50 km of optical fiber. Phys Rev Lett 2004; 93: 180502.


\bibitem{robust3} Graham TM, Bernstein HJ, Wei TC, Junge M, Kwiat PG. Superdense teleportation using hyperentangled
photons. Nat Commun 2015; 6: 7185.

\bibitem{Hu2018}
Hu XM, Guo Y, Liu BH, Huang YF, Li CF, Guo GC. Beating the channel capacity limit for superdense coding with entangled ququarts. Sci Adv 2018; 4: eaat9304.

\bibitem{Ecker2021}
Ecker S, Sohr P, Bulla L, Ursin R, Bohmann M. Remotely establishing polarization entanglement over noisy polarization channels. arXiv:2110.04159.

\end{thebibliography}
\end{document}